\documentclass[namedreferences,color]{solarphysics}
\usepackage[optionalrh]{spr-sola-addons} 
\usepackage{graphicx}        
\usepackage{color}           
\usepackage{url}             

\begin{document}

\begin{article}

\begin{opening}

\title{Atmosphere dynamics of the active region NOAA 11024 \\ {\it Solar Physics}}

\author{N. N.~\surname{Kondrashova}$^{1}$\sep
        M. N.~\surname{Pasechnik}$^{1}$\sep
        S.N.~\surname{Chornogor}$^{1}$\sep
        E. V.~\surname{Khomenko}$^{2,3}$
       }
\runningauthor{Kondrashova et al.} \runningtitle{Atmosphere dynamics}

\institute{$^{1}$ Main Astronomical Observatory, National Academy
of Sciences of Ukraine, 27 Akademika Zabolotnoho St., 03680, Kyiv,
Ukraine; email: \url{kondr@mao.kiev.ua} email: \url{rita@mao.kiev.ua} email: \url{chornog@mao.kiev.ua}\\
$^{2}$ Instituto de Astrofisica de Canarias (IAC), E-38200 La
Laguna, Tenerife, Spain; email: \url{khomenko@iac.es}\\
$^{3}$ Departamento de Astrof\'{\i}sica, Universidad de La Laguna,
38205, La Laguna, Tenerife, Spain}

\begin{abstract}
We present results of the study of chromospheric and photospheric
line-of-sight velocity fields in the young active region NOAA
11024. Multi-layer, multi-wavelength observational data were used
for the analysis of the emerging flux in this active region.
Spectropolarimetric observations were carried out with the
telescope THEMIS on Tenerife (Canary Islands) on 4 July 2009. In
addition, space-borne data from SOHO/MDI, STEREO and GOES were
also considered. The combination of data from
ground- and space-based telescopes allowed us to study the
dynamics of the lower atmosphere of the active region with high
spatial, spectral, and temporal resolutions. THEMIS spectra show
strong temporal variations of the velocity in the chromosphere and
photosphere for different activity features: two pores, active and
quiet plage regions, and two surges. The range of variations of
the chromospheric line-of-sight velocity at the heights of
formation of the H$\alpha$ core was extremely large. Both upward
and downward motions were observed in these layers. In particular,
a surge with upward velocities up to -73 km s$^{-1}$ were
detected. In the photosphere, predominantly upward motions were
found, varying from -3.1 km s$^{-1}$ upflows to 1.4 km s$^{-1}$
downflows in different structures. The velocity variations at
different levels in the lower atmosphere are compatible with
magnetic flux emergence.
\end{abstract}
\end{opening}

\section{Introduction}
     \label{S-Introduction}

The transition phase between Solar Cycles 23 and 24 was unusually
long and had a lot of peculiarities, including unusually low
levels of sunspot, flare, and geomagnetic activity. The Sun was
very quiet in 2009 -- 260 days without sunspots. Such low activity
in the transition period was not predicted and became a surprise.
Observational data may contribute to a better understanding of the
physical processes at this period and the beginning of the new
solar cycle.

In this work we study the line-of-sight velocity field in the
lower atmosphere of the active region NOAA 11024 observed on the
solar disk from 29 June to 11 July 2009. NOAA 11024 was the first
active region belonging to the new cycle that emerged in the
southern hemisphere of the Sun. It appeared at a latitude of about
-27 degrees. This is consistent with the butterfly diagram,
\textit{i.e.} that at the start of the cycle active regions began
to form at high latitudes on the Sun. This emerging active region
represents a very interesting case for a comprehensive study.

Some observational data of this region were collected and analyzed
by Brosius and Holman (2010), Schlichenmaier, Rezaei, and Bello
Gonzalez (2012), Valori \textit{et al.} (2012), Schad \textit{et al.}
(2011), Sylwester \textit{et al.} (2011), Engell \textit{et al.}
(2011). Schlichenmaier, Rezaei, and Bello Gonzalez (2012) have
investigated in detail the process of formation of a
well-developed penumbra at the large leading spot and have
obtained the line-of-sight (LOS) velocity maps in four instances
of 4 July 2009 using observations from German VTT and SOHO/MDI.
Schad \textit{et al.} (2011) have observed strong downflows with
supersonic velocities above the central part of the active region
and upflows near the leading spot on 7 July 2009. Valori
\textit{et al.} (2012) have studied the flux emergence process
between 29 June and 7 July 2009 by means of multi-wavelength
observations and nonlinear force-free extrapolation. Using high
spatial resolution observations they reviewed small-scale flux
emergence in the context of the global evolution of the active
region.

It is known that plasma motions play a significant role in the
processes of formation and development of active regions. The
conditions of new magnetic flux emergence determine the types of
the subsequent magnetic reconnections and the eruptive
instability in the active region. Numerous observations of young
active regions have been discussed in the literature. Their
dynamical properties have been studied by Howard (1971), Lites,
Skumanich, and Martinez Pillet (1998), Strous and Zwaan (1999),
Kubo, Shimizu, and Lites (2003), Kozu \textit{et al.} (2006),
Cheung, Sch\"ussler, and Moreno-Insertis (2006), Grigor'ev,
Ermakova, and Khlystova (2007, 2009), and others. Aznar Cuadrado,
Solanki, and Lagg (2005), Lagg \textit{et al.} (2007), and Schad
\textit{et al.} (2011) studied by means of spectropolarimetry in
He\,{\sc I} 1083 nm line, chromospheric gas flows in active
regions and concluded that further multi-line studies of
chromospheric flows are necessary. To form a more complete picture
of the physical processes, it is also important to study the
chromospheric and photospheric motions simultaneously.

In a number of previous works, a close relationship is found
between magnetic flux emergence and activity in the chromosphere
and corona. Kurokawa and Kawai (1993), Yokoyama and Shibata
(1996), Schmieder \textit{et al.} (2004), Li \textit{et al.}
(2007), and others emphasized the role of the magnetic
reconnection between emerging and pre-existing magnetic fields in
transient phenomena, such as surges, flares, microflares, jets,
and Ellerman bombs. Emerging flux can cause enhanced emission in
the Ca\,{\sc II} H line, as well as heating and brightening of
coronal loops.

In this paper we study the chromospheric and photospheric velocity
field in the young active region NOAA 11024 using spectral data
from the French-Italian telescope THEMIS at the Teide Observatory
of the Instituto de Astrofisica de Canarias, obtained on 4 July
2009 from 9:30 UT to 9:50 UT. We also consider data from space
telescopes SOHO, STEREO and GOES for a better understanding of the
physical processes leading to the dynamical features described in
this work.

\section{Active Region NOAA 11024} 
      \label{S-general}

\begin{figure}    
    \centerline{\includegraphics[width=1.00\textwidth,clip=]{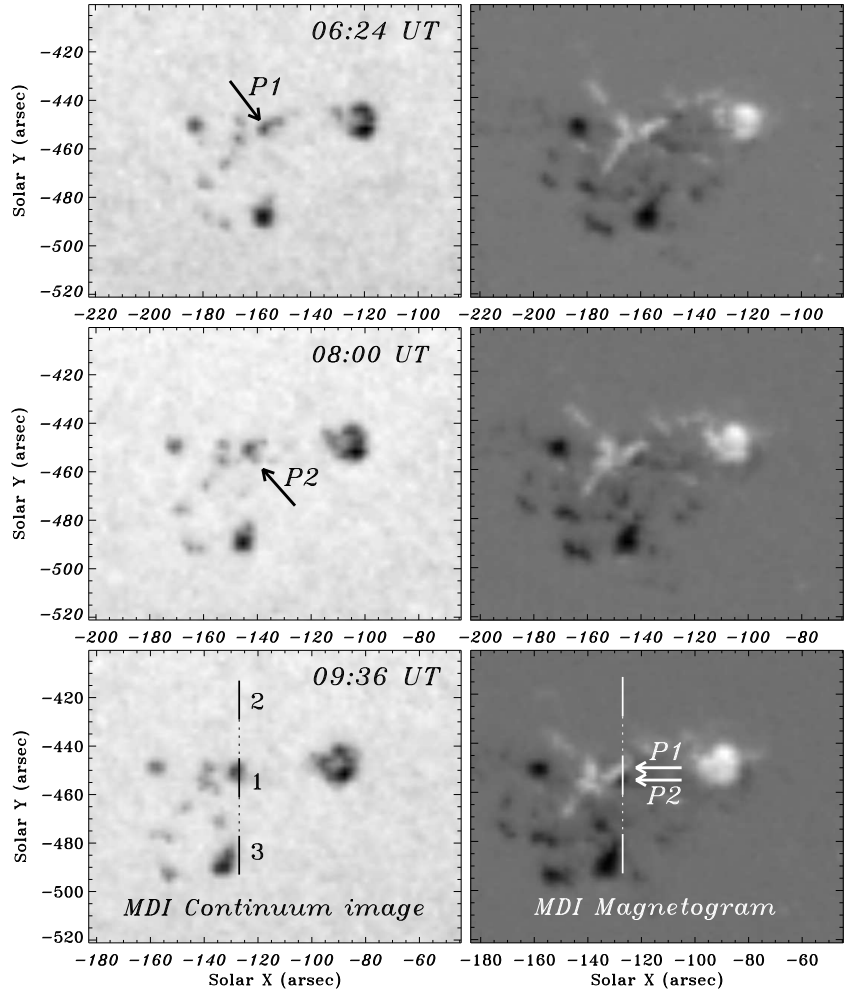}}
    \caption{SOHO/MDI continuum images (left panel) and MDI
    magnetograms (right panel) of the active region NOAA 11024 on 4
    July 2009 at 06:24, 08:00 and 09:36 UT. White and black colors
    show positive and negative polarities, respectively. P1 -- first
    and P2 -- second pores. The location of the THEMIS spectrograph slit is
    indicated with the line segments. Solid line (parts 1, 2, 3)
    denotes the parts in our field of view. Part 1 is the
    region we have investigated in detail in this paper. }
    \label{Fig1}
    \end{figure}

The young active region NOAA 11024 appeared on 29 June 2009 as a
rapidly growing facular region. It was the only active region on
the solar disk at that time. At the day of our observations at
THEMIS, on 4 July 2009, this active region was located near the
central meridian. It consisted of several spots and pores of
opposite polarities. Our observations took place during the phase
of fast rise of the activity in this active region (Valori
\textit{et al.}, 2012; Engell \textit{et al.}, 2011). A high surge
activity was observed in it, which is usually the first signature
of magnetic flux emergence (Kurokawa and Kawai, 1993).

We first review data and results from a number of multi-wavelength
observations of this active region. For a better understanding of
the evolution of the active region before and during our
spectroscopic observations at THEMIS, we have used magnetograms
and continuum images (see Figure 1) obtained with the Michelson
Doppler Imager (MDI; Scherrer \textit{et al.}, 1995) onboard the
SOHO spacecraft (Fleck, Domingo, and Poland, 1995). The spot group
was bipolar, with small opposite polarity patches indicating the
emergence of new magnetic flux.

Valori \textit{et al.} (2012) revealed the presence of serpentine
magnetic fields in this active region, where the serpentine fields
lines interacted with a pre-existing large-scale flux system.
Engell \textit{et al.} (2011) observed 137 flare-like/transient
events using SphinX, a full-disk-integrated spectrometer. Many
small X-ray bursts were recorded by GOES on 4 July 2009.

The time series of EUV images from STEREO and SOHO at 17.1 nm and
19.5 nm have allowed us to analyze the spatial and temporal
evolution of the upper atmosphere of the active region. The data
discussed here were obtained using EUVI, part of the \textit{ Sun
Earth Connection Coronal and Heliospheric Investigation} (SECCHI;
Howard \textit{et al.}, 2008) suite of instruments onboard the
STEREO-A spacecraft. All the images showed bright loop systems in
the corona and the transition region, which rapidly changed in
time. Figure 2 presents the set of the EUV images obtained by
STEREO-A at 17.1 nm for different times corresponding to our
spectral observations period. One can see the expansion of the
heated plasma along the loops at 09:31 -- 09:36 UT and 09:43:30 --
09:48:30 UT. Such motions of the heated plasma in the coronal
loops, most likely triggered by magnetic reconnection, were
observed by many authors. \v{S}vestka and Howard (1979) found that
the brightening in young coronal loops may be associated with the
emergence of new magnetic flux near their footpoints. Qiu
\textit{et al.} (1999) reported on the expansion of the
brightening along C\,{\sc IV} 155 nm loops in the transition
region, accompanied by motions along the H$\alpha$ loops in an
active region observed on 27 September 1998. The footpoints of the
loop system were located at regions of mixed magnetic polarity.
The authors suggested that the expansion was caused by slow,
small-scale magnetic reconnection at the foot regions of the loop
system.

\begin{figure}    
    \centerline{\includegraphics[width=1.20\textwidth,clip=]{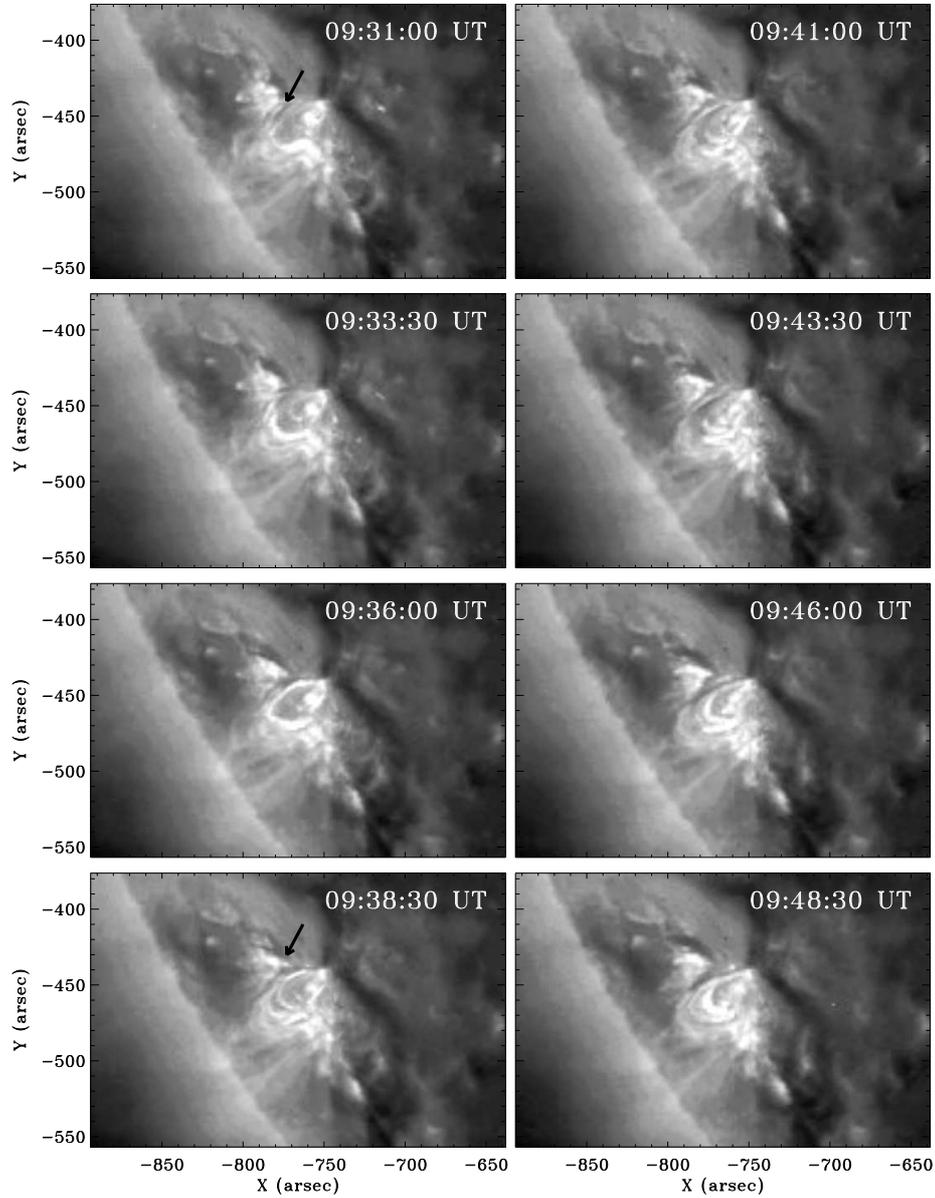}}
    \caption{EUV images (17.1 nm) of NOAA 11024 on 4 July 2009
    from STEREO-A/EUVI obtained during our observations.
    Surges are indicated by arrows.
    Time is given on each image.}
    \label{Fig3}
    \end{figure}

\section{Observations and data} 
      \label{S-general}

\begin{figure}    
    \centerline{\includegraphics[width=1.00\textwidth,clip=]{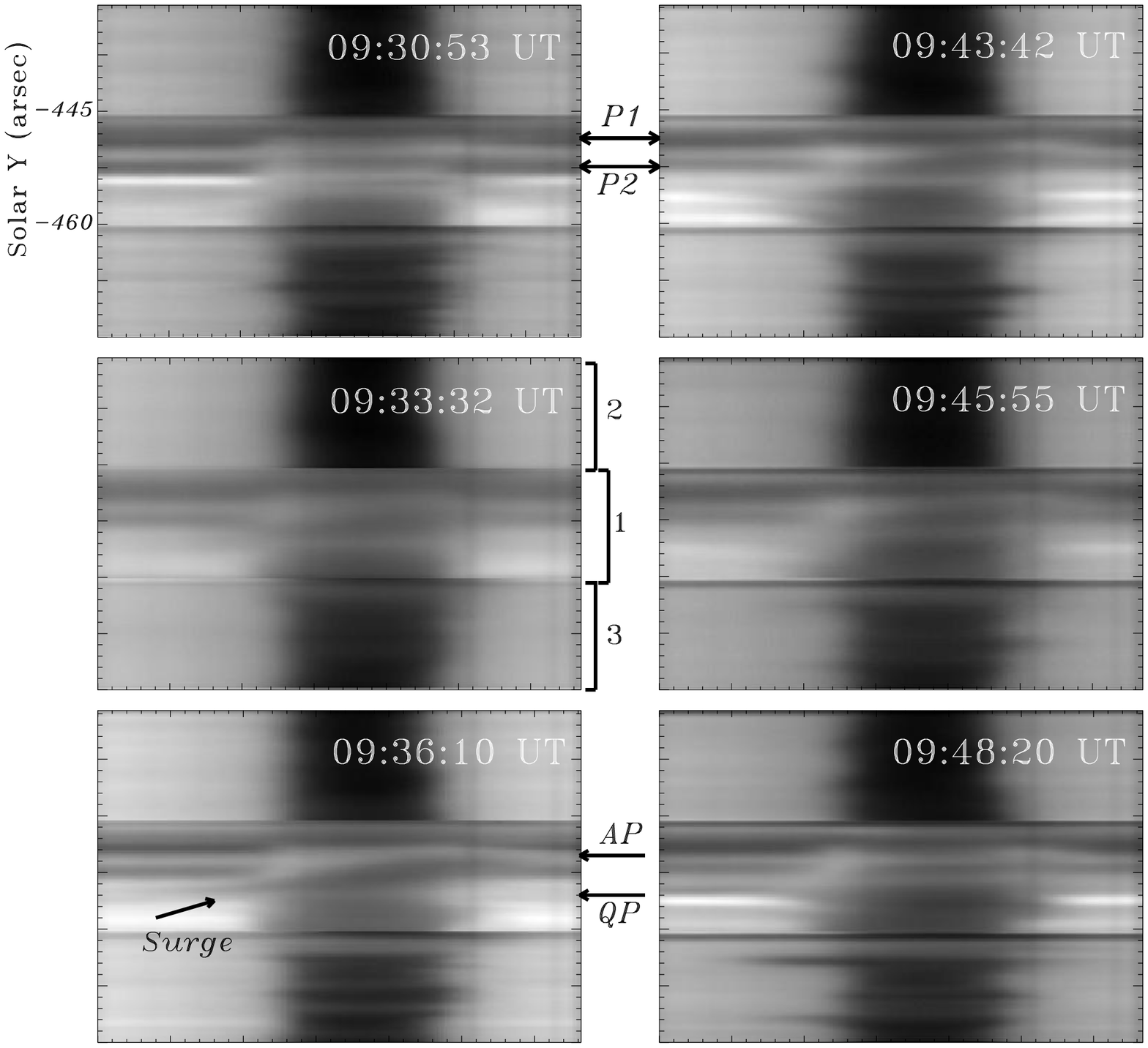}}
    \caption{H$\alpha$ spectra obtained during our observations.
    P1 and P2 -- pores, AP -- active plage, QP --
    quiet plage. 1, 2, 3 - the same as in Figure 1. }
    \label{Fig4}
    \end{figure}

Spectropolarimetric observations of the active region were carried
out with the French-Italian telescope THEMIS at the Teide
Observatory of the Instituto de Astrofisica de Canarias, on 4 July
2009. Here we are mostly interested in velocities and only
use Stokes $I$ profiles for our analysis.

Figure 1 (bottom panel) displays some MDI continuum images and
magnetograms around the time interval of our spectropolarimetric
THEMIS observations. Several small bipolar fragments were situated
between the main polarities. The location of the spectrograph slit
during the observations is given in Figure 1. The field of view of
THEMIS in polarimetric configuration is divided into three parts,
each of them about 16 arcsec long. The solid lines (parts 1, 2, 3)
denote the areas in the field of view and the dotted lines denote
the intervals between them. The part of the active region which we
investigated is labeled '1', and its useful length (after
reduction) was 14 arcsec (10 Mm).

The observed region included two pores and active and quiet
plages. H$\alpha$ spectra showed strong temporal changes in the
active plage, associated with the chromospheric surge.
There was no strong disturbance in the quiet plage region.
The location of the two pores, P1 and P2, is
indicated in Figure 1. In SOHO/MDI continuum images, at 06:24 UT
only the pore P1 (positive polarity) was observed. By 08:00 UT, P2
pore became visible, with the opposite (negative) polarity. New
emerging magnetic flux was apparent, as revealed by the comparison
of the MDI magnetograms obtained at 06:24 UT and 09:36 UT.

Several surges occurred in the studied active region during our
observations (Figures 2 and 3).

The pixel sampling of our observations was 0.2 arcsec, while the
actual spatial resolution, limited by the seeing effects, was
below 1 arcsec (the seeing conditions were exceptionally good
during the time of our observations). Fixed-slit time series were
taken, with a total of 400 spectra obtained between 09:30 UT and
09:50 UT. The time interval between individual spectra was
about 3 seconds, the exposure time was 0.12 seconds.

The complete data set included five spectral regions observed
simultaneously. In this work we use two of them, containing the
chromospheric line H$\alpha$ (central part) and the photospheric
lines Fe\,{\sc I} 630.15 nm, 630.25 nm, 630.35 nm, and the
Ti\,{\sc I} 630.38 nm. We have selected about one hundred spectra
of the best quality for this study. The time interval between the
selected consecutive spectra ranged from three to fifteen
seconds.

Figure 3 shows H$\alpha$ spectra obtained at approximately the
same time as the EUV images in Figure 2. The studied region is
marked by ``1'' and the location of other features is indicated.
It is seen that the spectra rapidly changed with time. Bright
emission regions, inclined bright and dark streaks indicate high
chromospheric activity. It should be noted that inclined streaks
are thought to indicate vortex-type chromospheric mass motions
associated with surges (Zirin, 1966).

\section{Results} 
      \label{S-Results}

From the observed Stokes $I$ profiles we have calculated values of
the line-of-sight velocities (on an absolute scale) in the
photosphere and chromosphere. The chromospheric velocities were
derived from the H$\alpha$ line core Doppler shifts using for
calibration the nearby telluric lines. The velocities in the
photosphere were calculated from the core shifts of the Fraunhofer
lines of Fe\,{\sc I} and Ti\,{\sc I} relative to the telluric line
present in this spectral window. Line-of-sight velocities were
measured by shifts of the line cores in the spectra relative to
their positions in the laboratory spectrum. All necessary
corrections were taken into account. The accuracy of the
chromospheric velocity is estimated to be about 0.3 km s$^{-1}$,
and that of the photospheric velocity about 0.1 km s$^{-1}$.

Sometimes a double component is clearly detected in the
H$\alpha$ profiles, with two clear minima. In these cases, the
position of both minima has been measured.

\begin{figure}    
    \centerline{\includegraphics[width=1.0\textwidth,clip=]{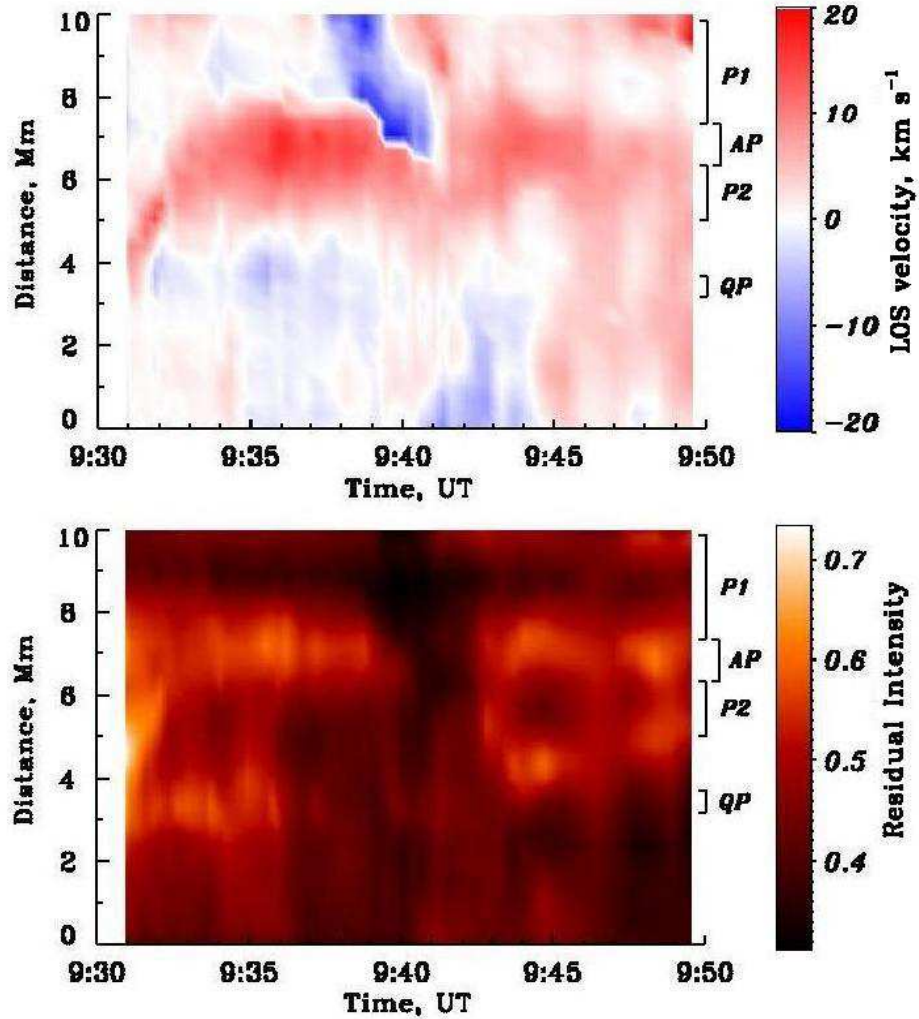}}
    \caption{Top: Temporal evolution of LOS chromospheric velocities in the active
    region along the spectrograph slit calculated from the H$\alpha$
    line core Doppler shifts. Positive velocities indicate downflows.
    The error is 0.3 km s$^{-1}$. P1 and P2 -- pores, AP --
    active plage, QP -- quiet plage. Bottom: Temporal variation of the
    residual intensity along the spectrograph slit at the center of the
    H$\alpha$ line.}
    \label{Fig5}
    \end{figure}

\subsection{Chromospheric line-of-sight velocities in the active region} 
  \label{S-Chromo-Velocities}

Figure 4 shows the temporal evolution of line-of-sight
chromospheric velocities and H$\alpha$ line core residual
intensities in the active region along the spectrograph slit.

\begin{figure}    
    \centerline{\hspace*{-0.12\textwidth}
    \includegraphics[width=0.85\textwidth,clip=]{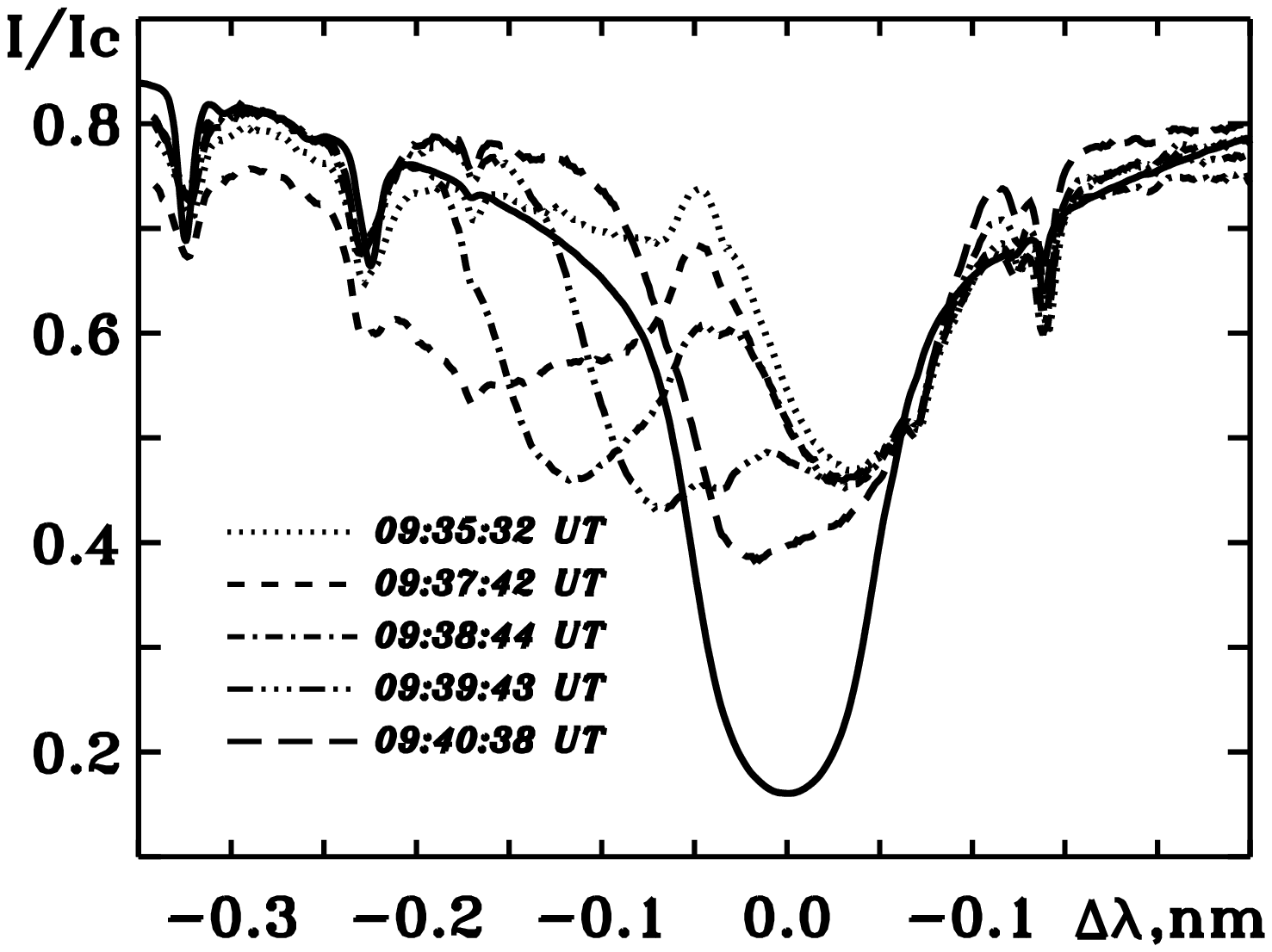}
    \hspace*{-0.03\textwidth}}
    \vspace{-0.35\textwidth}   
    \vspace{0.35\textwidth}    

    \centerline{\hspace*{-0.01\textwidth}
    \includegraphics[width=1.0\textwidth,clip=]{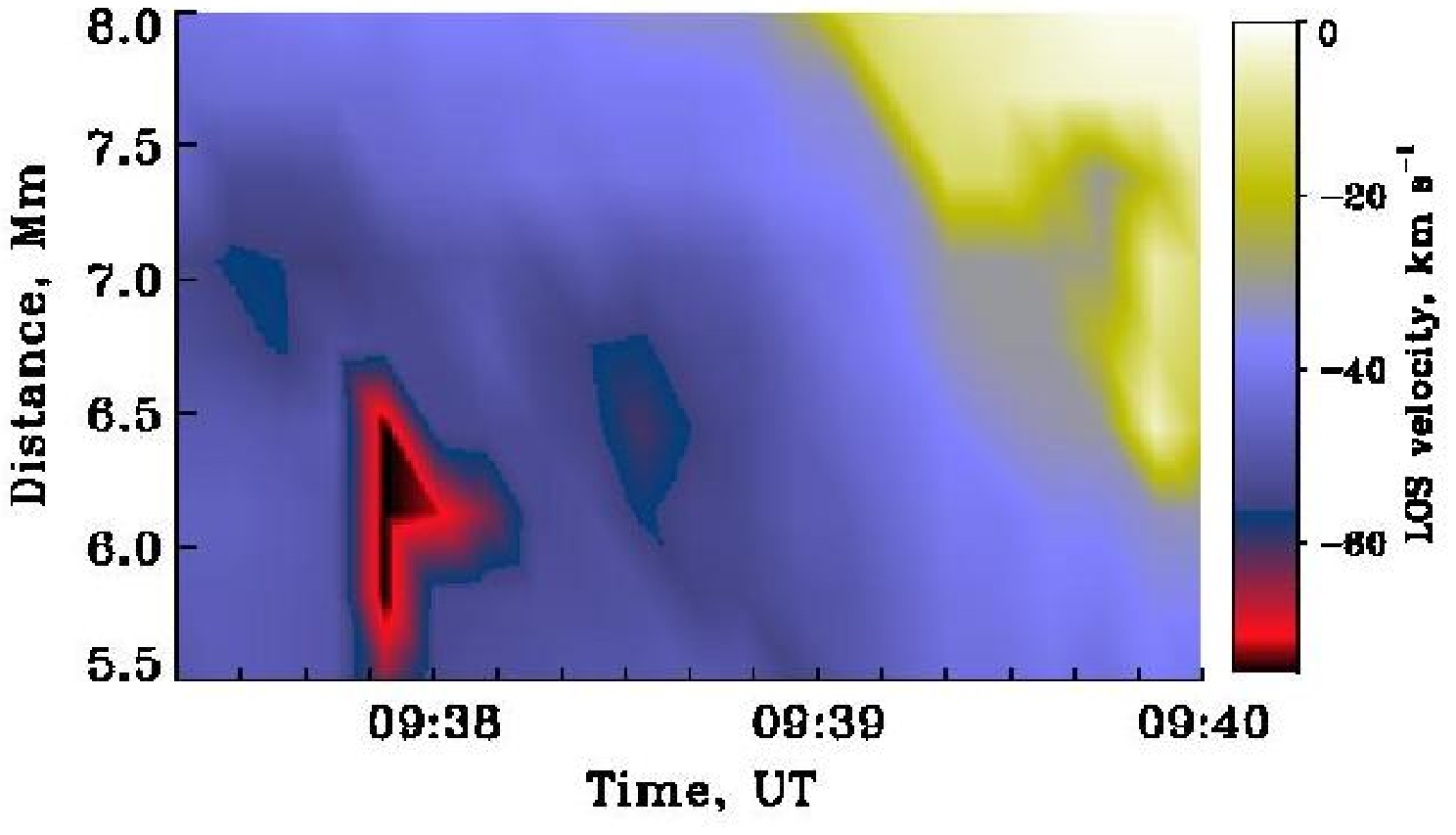}
    \hspace*{-0.03\textwidth}}
    \vspace{-0.35\textwidth}   
    \vspace{0.35\textwidth}    
    \caption{ Top: H$\alpha$ line profiles for different time moments
    corresponding to the second surge. For comparison, the quiet-Sun
    H$\alpha$ line profile from the atlas by Delbouille, Roland, and
    Neven (1973) is also plotted (solid line). Bottom:
    line-of-sight velocities during this H$\alpha$
    surge measured by the blue-shifted component of the line profiles.}
    \label{Fig6}
    \end{figure}

One can see that the chromospheric velocity field of the active
region was rather dynamic. During the first 15 min of the
observations both up- and downflows were detected. The last five
minutes, though, downflows were present over the whole region.
There are also differences in the velocity patterns between the
different active structures: pores, plage regions and surges.
Above the pore P1, the velocities and their changes during
observations were generally small (-1 -- 2 km s$^{-1}$) while at
the edges of the pore downflows were found up to 20 km s$^{-1}$.
Strong upflows between 09:37 and 09:41 UT are due to a surge, which
will be discussed below.

Above the pore P2, downflows were observed in the chromosphere.
The velocities were in the range of 1 km s$^{-1}$ on one side of
the pore, reaching 15 km s$^{-1}$ on the other side.

Plage regions also show strong velocity variations. In the active
plage during the first minute of the observations there was upflow
with the velocity, varying in time from -4 km s$^{-1}$ to about 0
km s$^{-1}$. Then the direction of motion changed and downflows
continued till the end of the observations.

Interestingly, a bright narrow band in both wings of the H$\alpha$
line was observed near the second pore (Figure 3, top left panel)
at the beginning of the observations. This region is seen in
Figure 4 (bottom panel) as a small bright region (at the location
4--5 Mm and time 09:31--09:32 UT) with increasing downward
velocity (top panel). In a minute this bright region moved in the
direction toward the active plage, and its brightness decreased
(Figure 4, bottom panel). This was accompanied by an increase in
the downward velocity from 6 km s$^{-1}$ to 14 km s$^{-1}$ (Figure
4, top panel).

In the quiet plage both up- and downflows, with a small velocities
around 0 km s$^{-1}$, were observed up to 09:45 UT, followed by
downflows with larger velocity afterwards.

Chromospheric surges occurred in the studied active region during
our observations. Sharp variations in the chromospheric velocity
direction, from downflow to upflow and \textit{vice versa}, were
revealed during these surges at about 09:34 -- 09:36 UT (first
surge) and at 09:37 -- 09:41 UT (second surge) at slit location
3.0 -- 4.2 Mm and 5.5 -- 10.0 Mm, respectively. In Figure 4 we can
see strong upflows in these places. The first surge was projected
over the region between the pore P2 and the quiet plage. The
second surge was projected over the region of the pore P1 and
active plage (upflow with velocities up to -20 km s$^{-1}$ in
Figure 4).

Figure 5 (top panel) shows the H$\alpha$ line profiles detected at
different time moments before and during the second surge. For
comparison, the quiet-Sun H$\alpha$ line profile from the atlas by
Delbouille, Roland, and Neven (1973) is given. It was corrected
for the limb darkening (cos $\theta$ = 0.87). The obtained line
profiles had two components -- main and blue-shifted. The
blue-shifted component suggests the appearance of a surge. We used
the shifts of this component to determine the surge velocity. It
should be noted that the intensity of this component sometimes
exceeded that of the main component. The bottom panel of Figure 5
displays the temporal variation of the velocity in the second
surge region.

All Doppler velocities were determined using the wavelength shift
of the local minimum of the H$\alpha$ line profiles. For Figure 4
it was the shift of the minimum of the main component, for Figure
5 (bottom) -- the shift of the minimum of the blue-shifted
component. At times close to the ejection maximum (in Figure 5 at
09:37:42 UT) this component had two minima, which means that the
surge was inhomogeneous, and consisted of several fragments. The
velocity peak ($\sim$ -73 km s$^{-1}$) was reached at 09:37:50 UT.

\subsection{Photospheric line-of-sight velocities in the active region} 
  \label{S-Photo-Velocities}

\begin{figure}    
    \centerline{\includegraphics[width=1.0\textwidth,clip=]{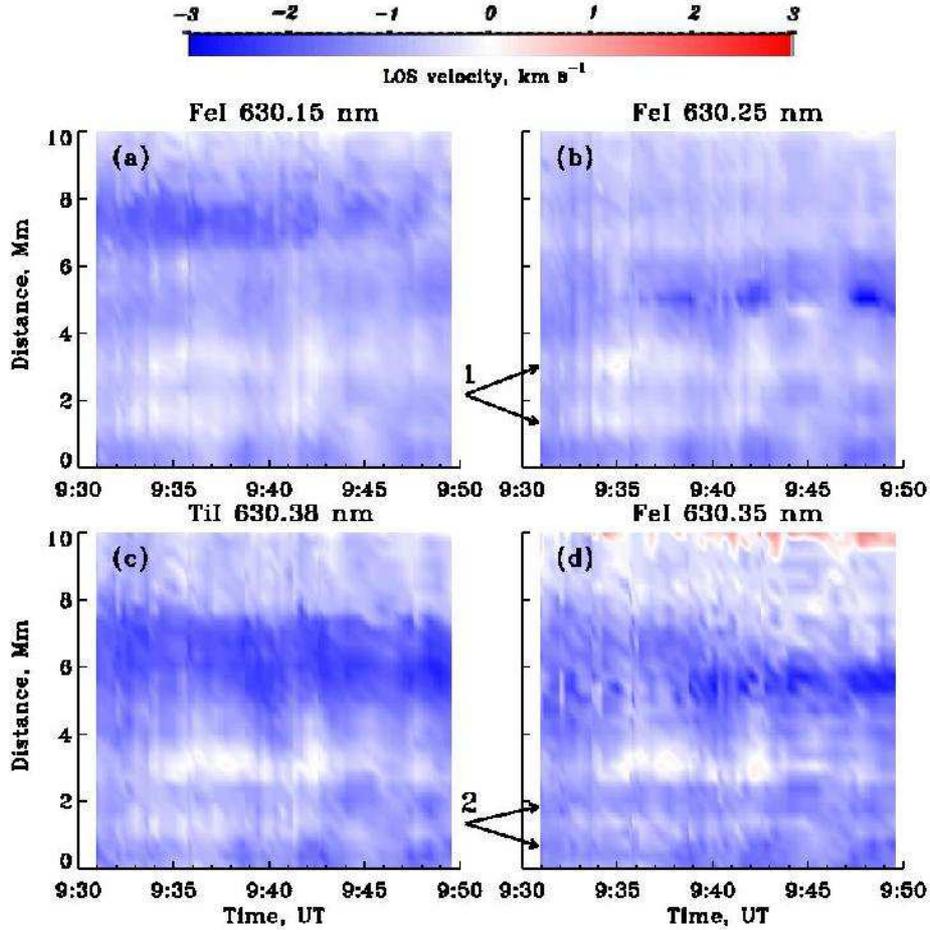}}
    \caption{Temporal changes of line-of-sight velocities in the
    active region along the spectrograph slit measured in four
    photospheric lines. The error is 0.1 km s$^{-1}$. Negative velocities
    indicate upflows. 1 -- intergranular lanes, 2 -- granules.}
    \label{F-simple}
    \end{figure}

Photospheric velocities were calculated using strong lines
Fe\,{\sc I} 630.15 nm, 630.25 nm and weak lines Fe\,{\sc I} 630.35
nm, Ti\,{\sc I} 630.38 nm. According to the atlas by Delbouille,
Roland, and Neven (1973), the central residual intensities of
these line profiles for the solar disk center in a quiet region
are 0.278, 0.344, 0.946, 0.916, respectively.

The plasma motion in the photospheric layers was different from
that in the chromosphere. Figure 6 shows the temporal variation of
the LOS velocity in the active region photosphere. Panels {\it a},
{\it b}, {\it c}, and {\it d} correspond to the lines Fe\,{\sc I}
630.15 nm, Fe\,{\sc I} 630.25 nm, Ti\,{\sc I} 630.38 nm, and
Fe\,{\sc I} 630.35 nm, formed in different layers of the
photosphere. From general considerations one may expect that, the
weaker lines Ti\,{\sc I} 630.38 nm and Fe\,{\sc I} 630.35 nm are
formed deeper and the stronger lines are formed higher, same as in
the quiet photosphere for the studied set of lines.

Mostly upflows were observed in all layers of the photosphere in
the region under consideration.

It is interesting to note the region with larger upflowing
velocities, whose location and width changed depending on the
height in the photosphere (Figure 6). The values of upflow
velocities changed ranged between -0.6 km s$^{-1}$ and -2.2 km
s$^{-1}$ for strong lines and between -1.3 and -2.7 km s$^{-1}$
for weak lines over time in this region.

The velocity variations in the central part of the first pore were
from -1 km s$^{-1}$ to -0.4 km s$^{-1}$. A large difference was
observed in the velocities at the edges of the pore. At one of the
boundaries, upflows of about -2 km s$^{-1}$ were detected over the
entire height of the photosphere. At the other boundary, the
upflow velocity was not large, only about -0.1 km s$^{-1}$. In the
formation region of the Fe I 630.35 nm line the flows had the
opposite direction, reaching 1.4 km s$^{-1}$. Schlichenmaier,
Rezaei, and Bello Gonzalez (2012) obtained a similar distribution
of the velocities for the leading spot of the same active region
during the formation of its penumbra.

The LOS velocities in the second pore derived
by the Fe\,{\sc I} 630.15 nm line (Figure 6, {\it a}) changed in time
from  -1 km s$^{-1}$ to -2 km s$^{-1}$. The velocities
for the Fe\,{\sc I} 630.25 nm line (Figure 6, {\it b}) were increasing
from -0.9 to -3 km s$^{-1}$.
In the formation region of the Ti\,{\sc I} 630.38 nm and
Fe\,{\sc I} 630.35 nm lines (lower layers of the photosphere)
the velocities were greater than in the formation region of
the Fe\,{\sc I} 630.15 nm and Fe\,{\sc I} 630.25 nm lines (upper layers)
during our observations (Figure 6, {\it c}, and {\it d}).

Granules and intergranular lanes can be identified in the lower
part of the velocity maps, where there were no active features
(pores and plages). In Figure 6 granules and intergranular lanes
are indicated by arrows (1 -- intergranular lanes
and 2 -- granules). Its LOS velocity changed
over time. Upflows were found predominantly both in granules and
intergranular lanes, with velocities -2 km s$^{-1}$ -- -0.6 km
s$^{-1}$ in granules and -0.9 -- 0.3 km s$^{-1}$ in the
intergranular lanes.

As a test of our result, and check the validity of a net upward
motion in the intergranular lanes of our active region, we have
considered the quiet sun region included in part ``2'' of the slit
(see Figure 1). In it, no apparent activity was present. In the
quiet photosphere, a typical granular pattern was obtained with
upward velocities of about -0.8 km s$^{-1}$ in granules and
downward velocities of about 1.5 km s$^{-1}$ in intergranular
lanes.

Thus, our results are compatible with a scenario where new
magnetic flux is appearing in the active region, superimposing an
upward velocity to the conventional overturning granular
convection.

\subsection{Comparison of chromospheric and photospheric line-of-sight velocities} 
  \label{S-Comparison}

\begin{figure}    
    \centerline{\includegraphics[width=1.0\textwidth,clip=]{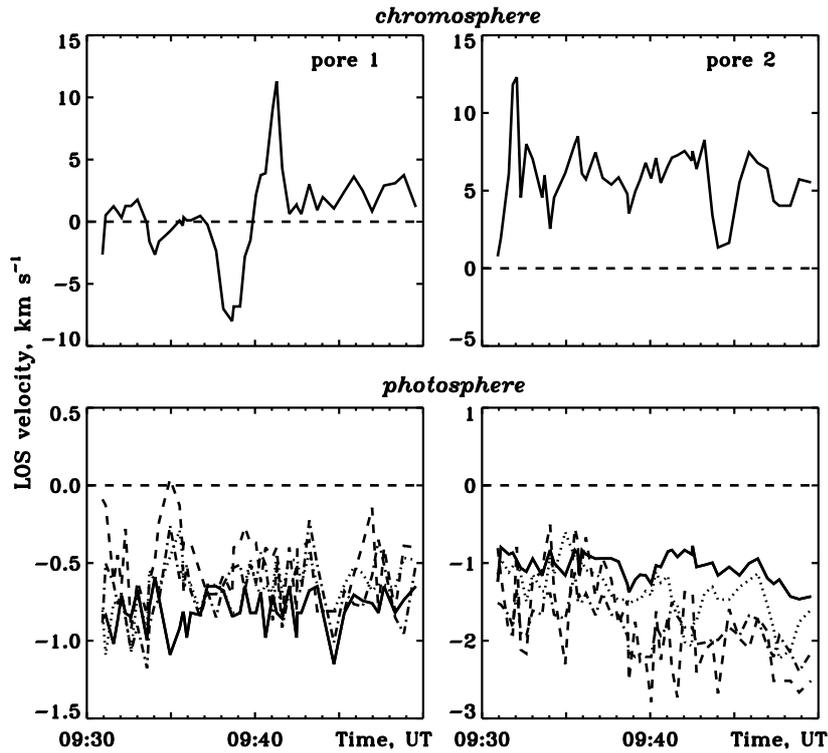}}
    \caption{Temporal variations of the chromospheric (top panels)
    and photospheric (bottom panels) velocities in the
    central part of the pore regions.
    On the bottom panels: solid line -- velocity for Fe\,{\sc I}
    630.15 nm, dotted line -- velocity for Fe\,{\sc I} 630.25 nm,
    dashed line -- velocity for Fe\,{\sc I} 630.35 nm, dotted-dashed
    line -- velocity for Ti\,{\sc I} 630.38 nm. }
    \label{F-simple}
    \end{figure}

To examine the relationship between the physical processes in the
chromosphere and photosphere we compared the velocities obtained
in these two layers in different features of the active region
(Figures 7 and 8).

Figure 7 shows significant differences between the chromospheric
and photospheric velocities in the pore regions. Preferentially
downflows were detected in the chromosphere above the pore P1.
In the photosphere upward velocity variations with a smaller
amplitude and mean value of -0.7 km s$^{-1}$ were found. The
largest changes in the chromospheric velocities above the pore P1
were associated with a surge overlaying it. The velocities
obtained from the profile main components (see section 4.1,
Figure 4, top panel) reached up to -8 km s$^{-1}$ at 09:38 UT
and after it up to 12 km s$^{-1}$.
It is seen that the chromospheric velocities above the
pore P2 were larger than above the first pore. The mean value of
the velocity was 7 km s$^{-1}$. In the photosphere of the second
pore, upflows were observed, with amplitude increasing over time
after the first surge up to about -1.5 -- -2.5 km s$^{-1}$. As in
the chromosphere, the photospheric velocities were larger,
compared to the first pore.

\begin{figure}    
    \centerline{\includegraphics[width=1.0\textwidth,clip=]{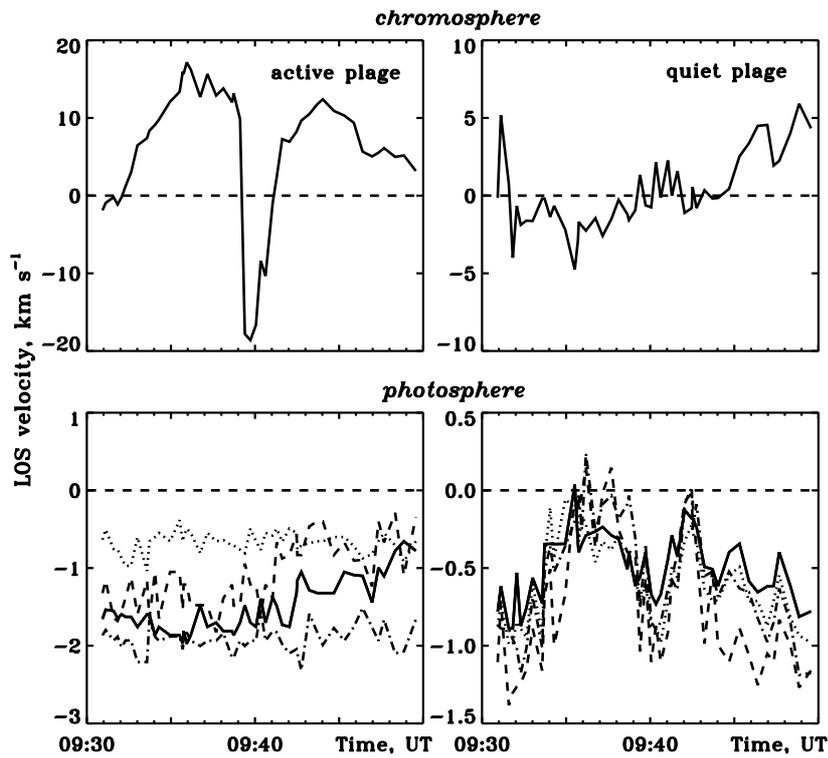}}
    \caption{Same as Fig.7, but in the central part of the active
    and quiet plage regions.}
    \label{F-simple}
    \end{figure}

In the active plage region, strong variations in the chromospheric
velocity were not accompanied by those at the photospheric levels
(Figure 8). Before and after the surge (09:37 UT  -- 09:41 UT)
strong downflows up to 18 km s$^{-1}$ were observed in the
chromosphere. Upflows from -2 km s$^{-1}$ to -0.5 km s$^{-1}$ at
different photospheric levels occurred.

In the quiet plage region, chromospheric downflows occurred at the
beginning of the observations. Then the motion direction rapidly
changed to upflows. After the first surge (peak at 09:35:40 UT)
the velocity direction changed again. The velocity increased with
time up to 6 km s$^{-1}$. In the photosphere the velocity
variations corresponded to typical photospheric oscillations.
Thus, the largest variations in the chromospheric LOS velocities
were observed in the surge regions. In the region of the new
magnetic flux emergence, in the surroundings of the pore P2,
velocities were largest both in the chromosphere and photosphere.

\section{Discussion and summary} 
      \label{S-Discussion and summary}

On the base of spectral observations with high temporal and
spatial resolution, we investigated the temporal variations of the
velocity field simultaneously in the chromosphere and at different
levels of the photosphere of the emerging active region NOAA
11024. On the day of our observations, there was a sharp increase
of its activity, evidenced by the rapidly changes in the EUV and
MDI images and H$\alpha$ spectra, as well as changes of the X-ray
flux. Many small X-ray bursts and chromospheric surges occurred in
the active region.

Line-of-sight velocities were determined in different features of
this young active region. The velocities were obtained in the
chromospheric H$\alpha$ line and four photospheric lines of
Fe\,{\sc I} 630.15 nm, 630.25 nm, 630.35 nm, and Ti\,{\sc I}
630.38 nm.

Our observations show rather strong temporal changes in the
chromospheric and photospheric velocities along a slice of the
active region of 10 Mm length, in an environment of emergence of
new magnetic flux. Chromospheric downflows up to about 18 km
s$^{-1}$ were observed in the vicinity of the magnetic flux
emergence region, in the active plage and P2 pore regions, while
LOS velocities in the quiet plage did not exceed $\pm 5$ km
s$^{-1}$.

Lagg \textit{et al.} (2007) have created the maps of the
chromospheric and photospheric LOS velocities in the active region
NOAA 9451 and found strong chromospheric downflows in the vicinity
of a young growing pore. The area of the downflow region was
increasing. Our observations show the expansion of the downflows
region as well. Lagg \textit{et al.} (2007) have analyzed in
detail the possible models of these downflows and interpreted them
being due to drainage of material along the legs of emerging
magnetic loops. As the pressure balance between the rising flux
tube and the surrounding atmosphere is changing, the hydrostatic
support of the the material in the upper part of the flux tube is
lost, and the material is draining back towards the deep layers.
There are other mechanisms explaining downward motions in the
loops, as, e.g. siphon flow models, model of the convective
collapse, etc. (\textit{e.g.}, Parker, 1978; Cargill and Priest,
1980; Grossmann-Doerth, Sch\"ussler, and Steiner, 1998; Takeuchi,
1999; Uitenbroek, Balasubramaniam, and Tritschler, 2006).

In this work, we also found upflows at all studied levels of the
photosphere. Ascending plasma seems to be present in the studied
slice of the active region. The largest upward velocities appeared
at the site of magnetic flux emergence, in the vicinity of a
polarity inversion line of the magnetic field.

Our results are compatible with earlier studies. Photospheric
upflows of -0.5 -- -2 km s$^{-1}$ are found in young, emerging
active regions (see, \textit{e.g.}, Howard, 1971; Lites,
Skumanich, and Martinez Pillet, 1998; Strous and Zwaan, 1999;
Kubo, Shimizu, and Lites, 2003; Cheung, Sch\"ussler, and
Moreno-Insertis, 2006; Kozu \textit{et al.}, 2006; Grigor'ev,
Ermakova, and Khlystova, 2007). Grigor'ev, Ermakova, and Khlystova
(2009) found that maximum velocities occurred in the polarity
inversion line of bipolar magnetic pairs. Numerical simulations
show that upflows must appear in regions of emerging magnetic flux
(\textit{e.g.}, Shibata \textit{et al.}, 1989; Fan, 2001; Magara
and Longcope, 2003; Archontis \textit{et al.}, 2004). Our
investigation confirms all these findings.

There are numerous simulations of magnetic flux emergence through
the photosphere into the corona. Example may be extended from the
initial 2-dimensional MHD models (\textit{e.g.}, Shibata
\textit{et al.}, 1989; Nozawa \textit{et al.}, 1992; Longcope,
Fisher, and Arendt, 1996; Emonet and Moreno-Insertis, 1998) to a
more sophisticated 3-dimensional ones (\textit{e.g.}, Matsumoto
\textit{et al.}, 1993; Fan, 2001; Magara and Longcope, 2003;
Abbett and Fisher, 2003; Archontis \textit{et al.}, 2004;
Archontis, 2010). These models explain many of the observed
features of emerging active regions and eruptive events. Our
observational data are in a broad qualitative agreement with the
above mentioned theoretical results.

Despite the recent progress, our complete understanding of how
magnetic field emerges from the convection zone into the corona is
still missing. Magnetic flux emergence from the convection zone
and formation of active regions is a complex physical process
affecting all layers of the solar atmosphere. In order to reach a
deeper understanding, it is very important to analyze multi-layer,
multi-wavelength spatially and temporally resolved observational
data, as those presented in this paper. Further progress is
expected with the coming era of the large-aperture ground-based
solar telescopes, as well as space missions.

\begin{acks}

We thank the referee for helpful suggestions for improvements of
this paper, the Editor Lidia van Driel-Gesztelyi and the Guest
Editor for constructive comments and corrections.
We would like to thank R. I. Kostyk for the codes for the
treatment of the observations.
We are grateful to the observer teams of the GOES,
STEREO and SOHO/MDI who have provide free access to their results.
We thank the support of the THEMIS team during the observations.
\end{acks}


\end{article}


\begin{thebibliography}{}

\bibitem[\protect\citeauthoryear{Abbett \& Fisher}{2003}]{Abbett03}
    Abbett, W.P., Fisher, G.H.: 2003,
    {\it Astrophys. J.} {\bf 582}, 475.

\bibitem[\protect\citeauthoryear{Archontis}{2010}]{Archontis10}
    Archontis, V.: 2010, {\it In K. Tsinganos, D. Hatzidimitriou, and
    T. Matsakos (Eds.): Adv. Hellenic Astron. IYA09}, {\bf ASP CS-424}, 3.

\bibitem[\protect\citeauthoryear{Archontis et al}{2004}]{Archontis04}
    Archontis, V., Moreno-Insertis, F., Galsgaard, K., Hood, A., O'Shea, E.: 2004,
    {\it Astron. Astrophys.} {\bf 426}, 1047.

\bibitem[\protect\citeauthoryear{Aznar Cuadrado et al}{2005}]{Cuadrado05}
    Aznar Cuadrado, R., Solanki, S.K., Lagg, A.: 2005,
    {\it in: Chromosphere and Coronal Magnetic Fields, edited by D.E. Innes,
    A. Lagg, S.K. Solanki, and D. Danesy}, (ESA Publication Division),
    {\bf ESA SP-596}, 49.

\bibitem[\protect\citeauthoryear{Brosius et al}{2010}]{Brosius10}
    Brosius, J.W., Holman, G.D.: 2010
    {\it Astrophys. J.} {\bf 720}, 1472.

\bibitem[\protect\citeauthoryear{Cargill \& Priest}{1980}]{Cargill80}
    Cargill, P.J., Priest, E.R.: 1980,
    {\it Solar Phys.} {\bf 65}, 251.

\bibitem[\protect\citeauthoryear{Cheung et al}{2006}]{Cheung06}
    Cheung, M.C.M., Sch\"ussler, M., Moreno-Insertis, F.:
    2006, {\bf ASPC 354}, 97.

\bibitem[\protect\citeauthoryear{Delbouille et al}{1973}]{Delbouille73}
    Delbouille, L., Roland, G., Neven L.: 1973
    {\it Photometric Atlas of the Solar Spectrum from
    3000 to 10000} Liege: Institut d'Astrophysique.

\bibitem[\protect\citeauthoryear{Emonet \& Moreno-Insertis}{1998}]{Emonet98}
    Emonet, T., Moreno-Insertis, F. 1998,
    {\it Astrophys. J.} {\bf 492}, 804.

\bibitem[\protect\citeauthoryear{Engell et al}{2011}]{Engell11}
    Engell, A.J., Siarkowski, M., Gryciuk, M., Sylwester, Ja.,
    Sylwester, B., Golub, L., {\it et al.}: 2011,
    {\it Astrophys. J.} {\bf 726}, 12.

\bibitem[\protect\citeauthoryear{Fan}{2001}]{Fan01}
    Fan, Y.: 2001,
    {\it Astrophys. J.} {\bf 554}, L111.

\bibitem[\protect\citeauthoryear{Fleck et al}{1995}]{Fleck95}
    Fleck, B., Domingo, V., Poland, A.I.: 1995,
    {\it Solar Phys.} {\bf 162}, 1.

\bibitem[\protect\citeauthoryear{Grigor'ev et al}{2007}]{Grigor'ev07}
    Grigor'ev, V.M., Ermakova, L.V., Khlystova, A.I.: 2007,
    {\it Astronomy Letters} {\bf 33}, 766.

\bibitem[\protect\citeauthoryear{Grigor'ev et al}{2009}]{Grigor'ev09}
    Grigor'ev, V.M., Ermakova, L.V., Khlystova, A.I.: 2009,
    {\it Astronomy Reports} {\bf 53}, 869.

\bibitem[\protect\citeauthoryear{Grossmann-Doerth et al}{1998}]{Grossmann-Doerth98}
    Grossmann-Doerth, U., Sch\"ussler, M., Steiner, O.: 1998,
    {\it Astron. Astrophys.} {\bf 337}, 928.

\bibitem[\protect\citeauthoryear{Howard}{1971}]{Howard71}
    Howard, R.A.: 1971,
    {\it Solar Phys.} {\bf 16}, 21.

\bibitem[\protect\citeauthoryear{Howard et al}{2008}]{Howard08}
    Howard, R.A., Moses, J.D., Vourlidas, A., Newmark, J.S., Socker,
    D.G., Plunkett, S.P., {\it et al.}: 2008,
    {\it Space Sci. Rev.} {\bf 136}, 67.

\bibitem[\protect\citeauthoryear{Kozu et al}{2006}]{Kozu06}
    Kozu, H., Kitai, R., Brooks, D.H., Kurokawa, H., Yoshimura, K., Berger, T.: 2006,
    {\it Publ. Astron. Soc. Japan} {\bf 58}, 407.

\bibitem[\protect\citeauthoryear{Kubo et al}{2003}]{Kubo03}
    Kubo M., Shimizu, T., Lites, B.W.: 2003,
    {\it Astrophys. J.} {\bf 595}, 465.

\bibitem[\protect\citeauthoryear{Kurokawa \& Kawai}{1993}]{Kurokawa93}
    Kurokawa, H., Kawai, G.: 1993,
    {\it in: The magnetic and velocity fields of solar active
    regions. ASP Conference Series, edited by H. Zirin, G. Ai,
    and Wang H.} {\bf 46}, 507.

\bibitem[\protect\citeauthoryear{Lagg et al}{2007}]{Lagg07}
     Lagg, A., Woch, J., Solanki, S.K., Krupp, N.: 2007,
     {\it Astron. Astrophys.} {\bf 462}, 1147.

\bibitem[\protect\citeauthoryear{Li et al}{2007}]{Li07}
     Li, H., Sakurai, T., Ichimito, K., Suematsu, Y.,
     Tsuneta, S., Katsukawa, Yu., {\it et al.}: 2007,
     {\it Publ. Astron. Soc. Japan} {\bf 59}, 643.

\bibitem[\protect\citeauthoryear{Lites et al}{1998}]{Lites98}
    Lites, B.W., Skumanich, A., Martinez Pillet, V.: 1998,
    {\it Astron. Astrophys.} {\bf 333}, 1053.

\bibitem[\protect\citeauthoryear{Longcope et al}{1996}]{Longcope96}
    Longcope, D.W., Fisher, G.H., Arendt, S.: 1996,
    {\it Astrophys. J.} {\bf 464}, 999.

\bibitem[\protect\citeauthoryear{Magara}{2003}]{Magara03}
    Magara, T., Longcope, D.W.: 2003,
    {\it Astrophys. J.} {\bf 586}, 630.

\bibitem[\protect\citeauthoryear{Matsumoto et al}{1993}]{Matsumoto93}
    Matsumoto, R., Tajima, T., Shibata, K., Kaisig, M. 1993,
    {\it Astrophys. J.} {\bf 414}, 357.

\bibitem[\protect\citeauthoryear{Nozawa et al}{1992}]{Nozawa92}
    Nozawa, S., Shibata, K., Matsumoto, R., Sterling, A.C., Tajima,
    T., Uchida, Y., {\it et al.}: 1992,
    {\it Astrophys. J. Supl. Ser.}, {\bf 78}, 267.

\bibitem[\protect\citeauthoryear{Parker}{1978}]{Parker78}
    Parker, E.N.: 1978,
    {\it Astrophys. J.} {\bf 221}, 368.

\bibitem[\protect\citeauthoryear{Qiu et al}{1999}]{Qiu99}
    Qiu, J., Wang, H., Chae, J., Goode, Ph.R.: 1999,
    {\it Solar Phys.} {\bf 190}, 153.

\bibitem[\protect\citeauthoryear{Schad et al}{2011}]{Schad11}
    Schad, T.A., Jaeggli, S.A., Lin, H., Penn, M.J.: 2011,
    {\it In J.R. Kuhn, D.M. Harrington, H. Lin, S.V. Berdyugina,
    J. Trujillo-Bueno, S.L. Keil, and T. Rimmele (Eds.):
    Solar Polarization 6. San Francisco: Astron. Soc. Pacific.}
    {\bf ASP CS-437}, 483.

\bibitem[\protect\citeauthoryear{Scherrer et al}{1995}]{Scherrer95}
    Scherrer, P.H., Bogart, R.S., Bush, R.I., Hoeksema, J.T.,
    Kosovichev, A.G., Schou, J., {\it et al.}: 1995,
    {\it Solar Phys.} {\bf 162}, 129.

\bibitem[\protect\citeauthoryear{Schlichenmaier et al}{2012}]{Schlichenmaier12}
    Schlichenmaier, R., Rezaei, R., Bello Gonzalez, N.: 2012,
    {\it In L.R. Bellot Rubio, F. Reale, and M.Carlsson (Eds.):
    Hinode-4. San Francisco: Astron. Soc. Pacific.} {\bf ASP CS-455}, 61.

\bibitem[\protect\citeauthoryear{Schmieder et al}{2004}]{Schmieder04}
    Schmieder, B., Rust, D.M., Georgoulis, M.K., Demoulin, P., Bernasconi, P.N.: 2004,
    {\it Astrophys. J.} {\bf 601}, 530.

\bibitem[\protect\citeauthoryear{Shibata}{1989}]{Shibata89}
    Shibata, K., Tajima, T., Steinolfson, R.S., Matsumoto, R.: 1989,
    {\it Astrophys. J.} {\bf 345}, 584.

\bibitem[\protect\citeauthoryear{Strous \& Zwaan}{1999}]{Strous99}
    Strous, L.H., Zwaan, C.: 1999,
    {\it Astrophys. J.} {\bf 527}, 435.

\bibitem[\protect\citeauthoryear{Sylwester et al}{2011}]{Sylwester11}
    Sylwester, B., Sylwester, J., Siarkowski, M., Engell, A.J.,
    Kuzin, S.V.: 2011,
    {\it CEAB} {\bf 35}, 171.

\bibitem[\protect\citeauthoryear{Svestka \& Howard}{1979}]{Svestka79}
    \v{S}vestka, Z., Howard, R.: 1979,
    {\it Solar Phys.} {\bf 63}, 297.

\bibitem[\protect\citeauthoryear{Takeuchi}{1999}]{Takeuchi99}
    Takeuchi, A.: 1999,
    {\it Astrophys. J.} {\bf 522}, 518.

\bibitem[\protect\citeauthoryear{Uitenbroek et al}{2006}]{Uitenbroek06}
    Uitenbroek, H., Balasubramaniam, K.S., Tritschler, A.: 2006,
    {\it Astrophys. J.} {\bf 645}, 776.

\bibitem[\protect\citeauthoryear{Valori et al}{2011}]{Valori11}
    Valori, G., Green, L.M., D\'emoulin, P., Vargas Dominguez, S.,
    van Driel-Gesztelyi, L., Wallace, A., {\it et al.}: 2012,
    {\it Solar Phys.} {\bf 278}, 73.

\bibitem[\protect\citeauthoryear{Yokoyama \& Shibata}{1996}]{Yokoyama96}
    Yokoyama, T., Shibata, K.: 1996,
    {\it Publ. Astron. Soc. Japan} {\bf 48}, 353.

\bibitem[\protect\citeauthoryear{Zirin}{1966}]{Zirin66}
    Zirin, H.: 1966,
    {\it The Solar Atmosphere}, Blaisdell Publ. Co., Massachusetts, U.S.A.


\end{thebibliography}
\end{document}